\documentclass[a4paper,11pt]{article}
\usepackage{jheppub} 

\usepackage{amsmath}
\usepackage{graphicx}

\title{\boldmath Extra Dimensions, Dual Relation and the Cosmological Constant}

\author[a]{Yong-Chang HUANG}
\author[a]{and LiuJi LI}

\affiliation[a]{Institute of Theoretical Physics, Beijing University of Technology,\\
Pingleyuan 100, Chaoyang district, Beijing, 100124, China}

\emailAdd{ychuang@bjut.edu.cn}
\emailAdd{plilj79@gmail.com} 
\abstract{We consider a general n dimensional manifold, which is a direct product manifold of $M^4 \times M^{n-4}$ representing our universe and extra spatial dimensions. From Einstein-Hilbert action of the manifold, we deduce effective 4 dimensional Fridemann equations. Matching it with the normal Fridemann equations with the cosmological constant, we give an expression of the cosmological constant, and the expression under different conditions are discussed. Also according to a deduced relation, we reveal a connection between Hubble parameter and extra dimensions. In this regard,  we take into account all possible states of extra dimension which are consistent with current experimental result as well.}

\begin{document} 
\maketitle
\flushbottom

 \section{Introduction}

The idea of extra dimensions plays a central role in string theory,  and it also has important effect in particle physics phenomenology and theoretical cosmology. The common feature of theories with extra dimensions is that there are more spatial dimensions, compacted to tiny scale far beyond  current experiments, other than our well-known 4-dimensional spacetime.

We know that our universe is expanding after Hubble's great job, and the expansion state of our universe, accelerating or decelerating, attracts people's enormous attention, because it will decide the final fate of our universe. At the end  of last century, observational cosmological data from type Ia supernovae~\cite{Riess:1998cb,Perlmutter:1998np} strongly suggest that the expansion of the universe is accelerating, and an unknown force called dark energy is governing the accelerating expansion.  From theoretical point of view, dark energy can be considered as potential energy of a dynamic field, besides, there is also a nature candidate, the cosmological constant term $\Lambda$, in general relativity.

Works have been investigated on phenomenal models of the dynamic field.  In this paper, however, we would like to constrain our attention on issue of the cosmological constant term. Without need introducing any kind of dynamic field is an advantage for the cosmological constant, but we must answer what the cosmological constant is. From the viewpoint of  quantum field theory, the cosmological constant is the vacuum energy of all kinds of field. While, if that is the case,  there is a conflict between observational results of the cosmological constant and its outcome of theoretical calculation. Observational results ~\cite{Ade:2013zuv, Lahav:2014vza} show that its value is very small and is the same degree as matter density of universe. On the other hand, as pointed out long time ago, theoretical calculation~\cite{Weinberg:1988cp, Abbott:1988nx} of vacuum energy give a very large number to it, and the magnitude of theoretical prediction is  larger than the observational result by a factor of $10^{120}$. Though supersymmetry requires a exact cancellation of the vacuum energy between boson fields and fermion fields,  supersymmetry broken at low energy has closed the door on using supersymmetry to explain why the cosmological constant is so small. In this paper, we would like to consider the idea of extra dimensions,  we want to find out if there is any potential explanation of the cosmological constant from aspect of extra dimensions. 

There are many kind of theories with extra dimensions. Theories with one extra dimension are totally different with theories with six extra dimensions. Even with the same number, various metric of extra dimension could come up with distinct physics.  For generality of discussion, we will not put any constrain on extra dimension, and we only assume that there are $n-4$ extra dimensions and  we do not give any detail of the metric of extra dimensions.  We will follow this belief,  and we do not add any rules on formula of the metric, until we could not continue our discussion without new regulation. 
 
The paper is organised as follow. At section 2, we introduce a n dimensional manifold, and our 4 dimensional universe is its submanifold. Considering a given metric of the manifold, we arrive at its Einstein-Hilbert action at first, and then deduce 4-dimensional effective Einstein equation and Friedmann equations from the action; We discover a dual relation between scale factor of our universe and scale factor of extra dimensions at section 3, also we point an expression of the cosmological constant out in terms of contribution of extra dimensions; Section 4  discuss relation between the cosmological constant and extra dimensions, in addition, we give a relation of Hubble parameter and scale factor $B(t)$ of extra dimensions. At final part of the paper, we study the motion state of extra dimensions.

\section{Cosmology with extra dimensions}

Our universe with extra spatial dimensions can be described by a general n-dimensional spacetime manifold, and the most simple case is a direct-product manifold $M^4\times M^{n-4}$, where $M^4$ is a 4-dimensional spacetime representing our universe, and $M^{n-4}$ is a ($n-4)$-dimensional space manifold. For this n-dimensional manifold, we take its infinitesimal line element square as 
\begin{equation}\label{metric}
ds^2=g_{\mu\nu}dx^{\mu}dx^{\nu}+g_{ab} (t,x^a)dx^adx^b,\quad \mu,\nu=0,1,2,3;\quad a,b=1,2,...,n-4,
\end{equation} 
where $g_{\mu \nu } =dt^2-a^2(t)\left[\frac{dr^2}{1-kr^2}+r^2d\theta^2 +r^2 sin^2 \theta d\phi \right]$ is the Robertson-Walker metrics of our universe, $g_{ab}$ is the metric of extra spatial manifold. In principle, $g_{ab}$ could be a function of $x^{\mu}$ and $x^a$, but it increase only difficult of calculation. To extract the key information from extra dimension, we here consider that $g_{ab}$ is function of only time $t$ and coordinate $x^a$ of extra dimensions. In addition, we merely have condition $n\ge 5$, and we do not require $n\le 11$, even though we know already the total dimensional number of M-theory is 11, because our consideration has nothing to do with string theory. In the field theory language, Einstein-Hilbert action of a manifold is only related with its metric and Ricci scalar. The total Ricci scalar of the n-dimensional spacetime manifold can be given by
\begin{equation}
\label{totricci} 
\mathcal{R}_{total} =g^{AB}R_{AB} =g^{\mu \nu }R^\alpha _{\mu
\alpha \nu } +g^{ab}R^\alpha _{a\alpha b} +g^{\mu \nu }R^c_{\mu c\nu
} +g^{ab}R^c_{acb} =\mathcal{R}+\mathcal{R}'+\mathcal{R}_{cross}  ,
\end{equation}
where
$ \mathcal{R}=g^{\mu \nu }(\Gamma ^\alpha _{\mu \nu ,\alpha } -\Gamma ^\alpha
_{\mu \alpha ,\nu } +\Gamma ^\alpha _{\beta \alpha } \Gamma ^\beta
_{\mu \nu } -\Gamma ^\alpha _{\beta \nu } \Gamma ^\beta _{\mu \alpha
} )$ is the Ricci scalar of our 4-dimensional spacetime, $\mathcal{R}'=g^{ab}(\Gamma ^c_{ab,c} -\Gamma ^c_{ac,b} +\Gamma ^c_{dc} \Gamma^d_{ab} -\Gamma ^c_{db} \Gamma ^d_{ac} )$ is the Ricci scalar of manifold of extra spatial dimensions, and $\mathcal{R}_{cross} =g^{\mu \nu }(-\Gamma ^c_{\mu c,\nu } +\Gamma ^c_{\beta c} \Gamma ^\beta _{\mu \nu } -\Gamma ^c_{a\nu } \Gamma ^a_{\mu c} )  +g^{ab}(\Gamma ^\alpha _{ab,\alpha } +\Gamma ^\alpha_{\beta \alpha } \Gamma ^\beta _{ab} -\Gamma ^\alpha _{db} \Gamma^d_{a\alpha } +\Gamma ^c_{\beta c} \Gamma ^\beta _{ab} -\Gamma^c_{\beta b} \Gamma ^\beta _{ac} )$ is the cross term between the normal 4-dimensional space-time and extra manifold. Later, we will find this general decomposition expressions  of Ricci scalar curvature of the n-dimensional spacetime manifold very useful for calculation. The term $R_{cross}$ is always neglected in past works on extra dimensions, whereas, it represents the magnitude of the coupling of our universe and the extra dimensional manifold. 

With the help of metric (\ref{metric}) and Ricci scalar (\ref{totricci}), Einstein-Hilbert action of the n-dimensional space-time is given 
\begin{equation}
\label{actionn}
\mathcal{I}_n=\int {d^4xd^{n-4}y} \sqrt {-g_{total} } \mathcal{R}_{total}.
\end{equation}
Using variational principle on action (\ref{actionn}) with respect to $g_{\mu\nu}$, we obtain a 4-dimensional effective Einstein field equation in the n-dimensional spacetime as follows
\begin{equation}
\label{einsteing} 
R_{\mu \nu } -\frac{1}{2}g_{\mu \nu } \mathcal{R}-g_{\mu \nu } \frac{\mathcal{R}'+\mathcal{R}_{cross} }{2}=-C_{\mu \nu},
\end{equation}
where $C_{\mu\nu}$ has a complicated form
\begin{equation} \label{tcmn}
\begin{split} 
C_{\mu \nu}=& -\Gamma _{a\nu }^b \Gamma _{\mu b}^a +\Gamma_{\alpha b}^b \Gamma _{\mu \nu }^\alpha +\frac{g^{\alpha \beta}}{2}\Gamma _{\mu b}^b \{\nu ,\alpha \beta \}+\frac{g^{cd}}{2}\Gamma_{\mu b}^b \{\nu ,cd\} -\frac{g^{cd}}{2}\Gamma _{\mu d}^b \{\nu,cb\}\\
&+\frac{g^{cd}}{2}\partial _\mu \{\nu ,cd\}+\frac{g^{cd}}{2}\Gamma_{cd}^\alpha \{\nu ,\alpha \mu \}+\frac{g^{cd}}{2}\Gamma _{\mu\alpha }^\alpha \{\nu ,cd\}-\frac{g^{cd}}{2}\Gamma _{\mu c}^a \{\nu,ad\}\\
&{-\frac{1}{2}\partial _\mu (g^{cd}\{\nu,cd\})}+\frac{1}{2}\partial _\alpha (g^{cd}g_{\mu \nu } \Gamma_{cd}^\alpha ).
\end{split}
\end{equation}

As doing in general field theories, we should add a general matter field Lagrangian $\mathcal{L}_m$, including baryon matter and dark matter, into the gravitational action of (\ref{actionn}), then the Einstein field equation (\ref{einsteing}) should be revised
\begin{equation}
\label{einsteingm} 
R_{\mu \nu } -\frac{1}{2}g_{\mu \nu } \mathcal{R}-g_{\mu \nu }
\frac{\mathcal{R}'+\mathcal{R}_{cross} }{2}=-8\pi GT_{\mu \nu } -C_{\mu \nu}.
\end{equation}
Ricci scalar is easy to obtain by timing $g^{\mu\nu}$ on both sides of Einstein equation (\ref{einsteingm}) 
\begin{equation}
\label{ricciscalar}
\mathcal{R}=8\pi GT+ C -2(\mathcal{R}'+\mathcal{R}_{cross}) .
\end{equation}
where $C=g^{\mu\nu}C_{\mu\nu}$. Putting the Ricci scalar ($\ref{ricciscalar}$) back into the Einstein equation ($\ref{einsteingm}$), we reach another form of Einstein field equation.
\begin{equation}
\label{einequ} 
R_{\mu \nu} =-8\pi G(T_{\mu \nu } -\frac{1}{2} g_{\mu\nu } T)-(C_{\mu \nu} -\frac{1}{2}g_{\mu \nu } C)-\frac{1 }{2}g_{\mu \nu} (\mathcal{R}'+\mathcal{R}_{cross}).
\end{equation}

In general, matter in the universe can be considered  as perfect fluid whose energy-momentum tensor is $T^{\mu}_{\nu}=\text{diag}(\rho, -P,-P,-P)$. With the help of Robertson-Walker metric $g_{\mu\nu}$, the effective Friedmann equations of 4-dimensional space-time are deduced from Equation ($\ref{einequ}$) as follows
\begin{equation}
\label{eftfriedequ1}
3\frac{\ddot {a}}{a}+4\pi G(\rho +3P) = -\frac{1}{4}\partial _0 g^{ab}\partial
_0 g_{ab} +\frac{1}{4}g^{ab}\partial _0 g_{ab} g^{cd}\partial _0 g_{cd}
-\frac{\mathcal{R}'}{2},
\end{equation}
\begin{equation}
\label{eftfriedequ2}
\frac{\ddot {a}+2\dot {a}^2+2\kappa}{a^2} -4\pi G(\rho -P) = -\frac{\dot {a}}{a}g^{ab}\partial _0
g_{ab} -\frac{1}{2}\mathcal{R}'  .
\end{equation}

\section{Duality between Scale Factors}

Since we are looking up to a possible explanation of the cosmological constant term in 4 dimensional theory from a higher dimensional theory, intuitive way is to match equations of 4 dimensional theory with the effective 4 dimensional equations deduced from higher dimensional theory.  Friedmann equations in 4-dimensional spacetime with the cosmological constant term $\Lambda$, which can be obtained from action $\mathcal{I}_4=\int d^4x \sqrt{-g}(R+\mathcal{L}_m+\Lambda)$
\begin{equation}
\label{4dfriedequ1}
3\frac{\ddot {a}}{a}+4\pi G(\rho +3P)=-\Lambda \quad ,
\end{equation}
\begin{equation}
\label{4dfriedequ2}
\frac{{\ddot {a}+\dot {a}^2+2\kappa}}{a^2} -4\pi G(\rho -P)=-\Lambda .
\end{equation}
Comparing these two equations to equations (\ref{eftfriedequ1}) and (\ref{eftfriedequ2}), we find that right side of equations (\ref{eftfriedequ1}) and (\ref{eftfriedequ2}) are equal to each other. Then we come to have one equation
\begin{equation}
\label{relequ}
\frac{1}{4}\partial _0 g^{ab}\partial _0 g_{ab} -\frac{1}{4}g^{ab}\partial_0 g_{ab} g^{cd}\partial _0 g_{cd} =\frac{\dot {a}}{a}g^{ab}\partial _0g_{ab} .
\end{equation}

We have no any information of the metric of extra dimensions yet, but it does not matter, we can still give a rough estimate on different states of extra dimensions. If the metric of extra dimensions are static, which means that they are  independent on time $t$, thus $\partial_0 g_{ab}=\partial_0g^{ab}=0$, then both side of equation (\ref{relequ}) are obviously zero, so the equation holds in this condition. In that case,  equations (\ref{eftfriedequ1}), (\ref{eftfriedequ2}), (\ref{4dfriedequ1}) and (\ref{4dfriedequ2}) give a relation between the cosmological constant term $\Lambda$ and the curvature of extra dimensions $\mathcal{R}'$
\begin{equation}\label{ccexdi}
\Lambda=\frac{1}{2}\mathcal{R}'
\end{equation}
If there are static extra spatial dimensions, then the relation (\ref{ccexdi}) shows that the cosmological constant is nothing but the half of Ricci curvature of extra spatial dimensions. Value of the cosmological constant is measured by current observational experiments to be on the order of $10^{-52} m^{-2}$, thus, the curvature of extra spatial dimensions is two times of the cosmological constant. Ricci curvature of extra dimensions is small, and it means that extra spatial dimensions are very close to be flat, and moreover, that Ricci scalar of extra dimensions is not zero means refers to geometry of extra dimensions is neither flat nor maximal symmetric. 

On the other hand, extra spatial dimensions could be dynamic and are functions of time $t$. Considering simplify and following the same idea of comoving coordinate of R-W metric, we boldly suppose that the metric of extra spatial dimension can be written as
\begin{equation}
\label{mesd}
g_{ab} =-B^2(t)\tilde {g}_{ab} \quad ,
\end{equation}
where $B(t)$ is scale factor of extra dimensions, and $\tilde {g}_{ab} $ ( a, b = 1,2, {\ldots}, n-4 ) depends only on coordinates of extra spatial dimensions and we do not discuss its any kind of concrete formula, we thus have two following equations
\begin{equation}
\label{esdequ1}
g^{ab}\partial_0 g_{ab} =2(n-4)\frac{\dot {B}}{B} ,
\end{equation}
\begin{equation}
\label{esdequ2}
\partial _0 g^{ab}\partial _0 g_{ab} =-4(n-4)\frac{\dot {B}^2}{B^2}  ,
\end{equation}
Plugging equations (\ref{esdequ1}) and (\ref{esdequ2}) into equation (\ref{relequ}), then we achieve an equation 
\begin{equation}
\label{nrelequ}
(n-4) \frac{\dot{B}}{B} \left((n-3)\frac{\dot{B}}{B}+2 \frac{\dot{a}}{a}\right)=0
\end{equation}
$n-4>0$ is always valid, because our start point is existing of extra spatial dimensions. One possible solution of equation (\ref{nrelequ}) is from $\frac{\dot{B}}{B}=0$, it gives rise to scale factor of extra dimensions $B(t)$ does not depend on time, so it comes back to the conclusion (\ref{ccexdi}), as we have discussed already. So general solution of equation (\ref{nrelequ}) can be given by equation~\cite{Li:2005cpc}
\begin{equation}\label{dualequ}
\frac{\dot{a}}{a}=-\frac{n-3}{2}\frac{\dot{B}}{B}
\end{equation}
which solution points out a relation of scale factors $a(t)$ and $B(t)$
\begin{equation}
\label{relab}
a(t)=\alpha B^{-\frac{n-3}{2}}(t) ,
\end{equation}
where $\alpha$ is an arbitrary constant parameter which can not be determined in the paper at present. The relation (\ref{relab}) exhibits a dual relationship of scale factors between our 3-dimensional space and  extra spatial dimensions. As we know, in Robertson-Walker metric, scale factor $a(t)$ is proportional to physical distance by a constant number. Then the scale factor $a(t)$ is very large because our universe has a long historical expanding and Its physical size is to a large extent. With help of the relation (\ref{relab}), the scale factor of extra dimensions $B(t)$ must be very small, which means that extra dimensions are small too. This result is consistent with the general idea of Kluza-Klein theory and string theory in which extra dimension(s) is(are) compact to very small scale. 

There are other approaches to study scale factors of our universe and extra dimensions, and one similar relation~\cite{Ho:2010vv} $\frac{\dot{a}}{a}=-\frac{N}{3}\frac{\dot{B}}{B}$ has been founded. Though the relation is not exactly the same as equation (\ref{dualequ}), both of them indicate one information that expansion of our universe is fulfilled by the contraction of extra dimension.  

\section{Discussion}

We could have some solutions of scale factor $a(t)$ after we add conditions on Friedmann equations, then,  it is easy to get the relation between scale factor $B(t)$ and time $t$. It is well known that our universe has different stages in which different materials govern the evolution of universe.  We consider two different stages, radiation dominate and matter dominate,  of universe evolution, then we can get the solution of $a(t)$ from the simplified Friedmann equations (\ref{4dfriedequ1}) and (\ref{4dfriedequ2}). 

At the radiation dominate stage of universe, the solution of scale factor $a(t)$ is given by $ a(t)\propto t^{1/2}$, then with equation (\ref{hced}), the scale factor $B(t)$ of extra dimensions have one solution $B(t)=B_0 t^{\frac{-1}{n-3}} $, where $B_0$ is a dimensionless constant.  Another stage of universe is dominated by matter, which gives scale factor $a(t)\propto t^{\frac{2}{3}}$, as a result, scale factor of extra dimensions is $B(t)\propto t^{\frac{-4}{3n-9}}$. 

\begin{figure}\label{evol}
\includegraphics[width=0.65\textwidth]{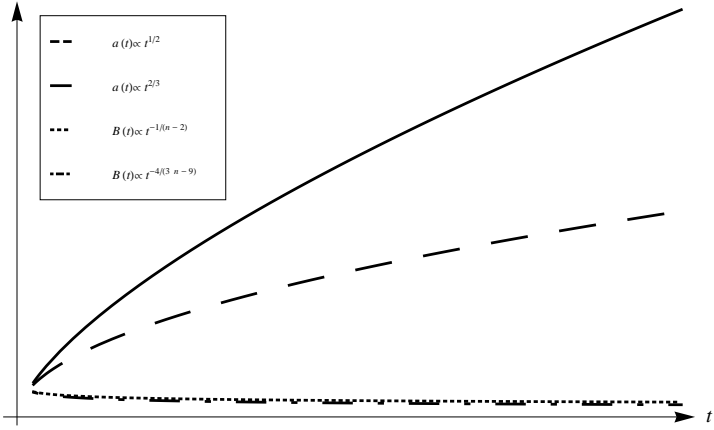}
\caption{evolution of scale factor a(t) and B(t) at different stage}
\end{figure}

We plot solutions of scale factors $a(t)$ and $B(t)$ at the different stages in the diagram \ref{evol}. It shows that evolution behaviour of our universe at different stages are obviously different if we normalize the solution of scale factor $a(t)$. While, it is not easy to distinguish normalized function $B(t)$ of scale factor of extra dimensions at different stages. The result is not surprising. If evolution of extra dimension varies too fast, there must be some effect on our universe. But experiments on testing variation of physical constant~\cite{Webb:2000mn} give negative results only, it indicates that evolution of extra dimension must be slow enough if there are dynamical extra dimensions.

According to the definition of Hubble parameter $H\equiv \frac{\dot{a}(t)}{a(t)}$ and equation (\ref{dualequ}), we have a new expression of Hubble parameter in terms of scale factor of extra dimensions
\begin{equation}
\label{hced}
H= - \frac{n-3}{2} \frac{\dot{B}}{B}
\end{equation}
This equation indicates that Hubble parameter is not only a symbol of expansion of our universe, but also a character of dynamic of extra dimension. The Hubble parameter is positive indicate expansion of our universe and shrinking of extra dimensions. The current Hubble's constant  $H_0$ is determined to be 67 by observational experimental data~\cite{Ade:2013zuv}, when we have number of extra spatial dimensions, we get immediately a parameter to represent dynamics of extra dimensions.

From equations (\ref{eftfriedequ2}),  (\ref{4dfriedequ2}) and (\ref{hced}), we have another relation equation
\begin{equation} \label{cosconst}
\Lambda=-(n-4)(n-3) \left(\frac{\dot{B}}{B} \right)^2 +\frac{1}{2} \mathcal{R}'=-\frac{4(n-4)}{n-3} H^2 +\frac{1}{2} \mathcal{R}'
\end{equation}

This equation shows that the cosmological constant is not a real constant, it relies strongly on shrinking rate and Ricci scalar of extra dimension. When a specific metric $\tilde{g}_{ab}$ of extra dimension is given, expression of the cosmological constant are settled down, then its evolution form can be studied,  and as well its effect on the evolution of our universe.  Moreover, Hubble parameter and the cosmological constant can be chosen as input parameter, their values can be given by updated experimental data, the left undetermined parameters are number $n$ and Ricci scalar  of extra dimensions $\mathcal{R}'$.  Current value of cosmological constant from Planck is given by $\Lambda \sim 10^{-52} m^{-2}$~\cite{Ade:2013zuv,Lahav:2014vza}, it is so small compare with the value of Hubble parameter which is given by $H^2 (t_0) \sim 1.4 \times 10^{-17} m^{-2}$ from Planck 2013 result~\cite{Ade:2013ktc},  here we can simply take cosmological constant to be zero, then we can have the  correspondence between  number $n$ of dimensions and Ricci scalar of extra dimensions. From table \ref{run}, we see that Ricci scalar of extra dimensions is dependent on total number of extra dimensions, and its current value has the same degree as the square of Hubble parameter.  Of course, the cosmological constant problem why it is so small is still there, but it is not the issue of this paper, we will not mention this point again.
\begin{table}[tb]
\caption{\label{rcn}
A table shows the relation of current Ricci scalar  of extra dimensions  and number of total dimensions.}
\begin{tabular}{| c |c|c|c|c|c|c|c|}
\hline
total dimensions number $n$ & 5 & 6 & 7 & 8 & 9 & 10 & 11 \\
\hline
Ricci scalar of extra dimensions ($10^{-17} m^{-2}$) &  5.6 &  7.5  & 8.4  &  8.9  & 9.3  &   9.6    & 9.8     \\
\hline
\end{tabular}
\end{table}

Besides, there is a deceleration parameter defined as $q=-\frac{\ddot{a}a}{\dot{a}^2}$, which is negative according to the results\cite{ Riess:1998cb, Perlmutter:1998np} of observations on distant type Ia supernovae.  It is clearly shown that  $\ddot{a}>0$ since scale factor $a$ and $\dot{a}^2$ are both positive.  By using equation (\ref{relab}), twice derivative on scale factor $a(t)$ with respect to time $t$ gives
\begin{equation}
\ddot{a}=\alpha (\frac{n-3}{2})(\frac{n-1}{2}) B^{-\frac{n+1}{2}} \dot{B}^2 + \alpha (-\frac{n-3}{2})B^{-\frac{n-1}{2}} \ddot{B}
\end{equation}
Then, condition $\ddot{a} >0$ indicates a constrain equation on $B$
\begin{equation}
\ddot{B}-\frac{n-1}{2} \frac{\dot{B}^2}{B} < 0
\end{equation}

We have known that $\dot{B}< 0 $ in accord with equation (\ref{relab}), however, it seems not possible to achieve any deceleration information of extra dimensions. Taking into consideration $\ddot{B}\le 0$, which means that shrinking speed of extra dimensions is becoming slow or keeping constant, the constrain equation always holds. Even $0<\ddot{B}< \frac{n-2}{2}\frac{\dot{B}^2}{B}$, the constrain equation is still to be valid. As for this point, we have to admit that we do not know how to extract more dynamic informations of extra dimensions at present. But it does not means that we will not get the informations forever, and it is possible to get a sign of extra dimensions after we spend more time to investigate on it.

\acknowledgments
The work is supported by National Natural Science Foundation of China grants No. 11275017 and No. 11173028. We thank Chiu Man HO for informing one similar relation of scale factor in ~\cite{Ho:2010vv}.  L.J.L thanks ITP at BJUT for kind hospitality.

\end{document}